\documentstyle[12pt]{article}
\newcommand{\be}{\begin{equation}}
\newcommand{\ee}{\end{equation}}
\newcommand{\PopisTab}[2]{{\bf Tab. #1: }{\footnotesize{#2}}}
\begin{document}
\frenchspacing
\title{Estimating the Fractal Dimension, $K_{2}$-entropy
 and the Predictability \\ of the Atmosphere \footnotetext{published in Czechoslovak Journal
 of Physics Vol.46 (1996) 293-328}}
\author{Ale\v s Raidl \\
{\small Department of Meteorology and Environment Protection, Charles University,} \\
{\small V~Hole\v sovi\v ck\' ach 2, 18200 Praha 8, Czech Republic,} \\
{\small e-mail: ar@kamet.troja.mff.cuni.cz}}
\date{February 8, 1995}
\maketitle
\thispagestyle{empty}
\newpage
\begin{abstract}
The series of mean daily temperature of air recorded over a~period of 215
years is used for analysing the dimensionality and the predictability of the
atmospheric system. The total number
of data points of the series is 78527. Other 37 versions of the original
series are generated, including ``seasonally adjusted'' data, a~smoothed series,
series without annual course, etc. \par
Modified methods of Grassberger \& Procaccia are applied.
A~procedure for selection of the ``meaningful'' scaling region is proposed.
Several scaling regions are revealed in the $\ln C(r)$ versus $\ln r$
diagram. The first one in the range of larger $\ln r$  has a~gradual slope
and the second one in the range of intermediate $\ln r$ has a~fast slope.
Other two regions are settled in the range of small $\ln r$. The results
lead us to claim that the series arises from the activity of at least two
subsystems. The first subsystem is low-dimensional ($d_f=1.6$) and it possesses
the potential predictability of several weeks. We suggest that this subsystem
is connected with
seasonal variability of weather. The second subsystem is high-dimensional
($d_f>17$) and its error-doubling time is about 4-7 days.\par
It is found that the predictability differs in dependence on season. The
predictability time for summer, winter and the entire year ($T_2 \approx
4.7$ days) is longer than for transition-seasons ($T_2 \approx 4.0$ days
for spring, $T_2 \approx 3.6$ days for autumn). \par
The role of random noise and the number of data points are discussed. It is
shown that a~15-year-long daily temperature series is not sufficient for
reliable estimations based on Grassberger \& Procaccia algorithms.
\end{abstract}
\section{Introduction}
One often observes phenomena which exhibit complicated nonperiodic behaviour
though they are controlled by strictly deterministic rules. Such processes
have also been recorded in the last twenty years in computational
simulations of many nonlinear models from various areas. The algorithms
did not contain any stochastic terms and/or numerical stability conditions
were not violated at the same time. Systems with similar behaviour were
already known in the last century, but the introduction of concept of the
strange attractor and the use of efficient computers have led to better
understanding of these phenomena. \par
1963 is regarded as the beginning of the theory of deterministic chaos which
studies such systems. In that year the well-known Lorenz's article
concerning some stage of atmospheric convection was published.
Lorenz [20] has designed system of three nonlinear differential
equations whose solutions exhibit a~chaotic evolution in time for
particular values of control parameters. The Lorenz system has provided
the first example of deterministic dissipative system with sensitive
dependence on initial conditions. Since then the deterministic chaos
has been detected
in many areas, for example in geophysics, chemistry,
biology, medicine and psychology, ecological systems, sociology, in rail
vehicle dynamics, etc. (see [22]).
\section{Dimension}
Dissipative systems with chaotic behaviour often possess a~strange
attractor. One of the principal characteristics which is used for
description of strange attractors is a~dimension. The dimension
reflects the complexity and strangeness of an attractor. Chaotic
attractors have a~noninteger dimension. Integer part of a~fractal
dimension $d_f$ plus one is the minimal number of independent variables
needed to describe the time evolution of the system. The maximal number of such
variables is $2d_f+1$ for a~simple system. There exists a~variety of
dimension definitions. The simplest one is the capacity $d_c$ of the
attractor $A$:
\be
d_{c}=\lim_{r \to 0} \frac{\ln M(r)}{\ln r}, \label{kapacita}
\ee
where $M(r)$ is the minimal number of $n$-dimensional cubes of side $r$ needed
to cover the attractor $A$. More complicated one is the Hausdorff dimension
$d_H$ (see [6], [7]).
Generally $d_c \geq d_H$ but it is conjectured that these dimensions
are the same for typical attractors. Their common value is called
the fractal dimension $d_f$ [22]. It must be noted
that roughly ten different fractal dimensions are used for a~description
of fractal sets [4]. \par
The above definition of the capacity is based on
metric properties of the attractor. Other definitions take into
account the frequency of visiting individual parts of the attractor
by the trajectory. An information dimension $d_1$ and a~correlation
dimension $d_2$ belong among most used dimensions. Both
dimensions are easily obtained from a~generalized dimension of order $q$
\be
d_q=-\lim_{r \to 0} \frac{1}{1-q} \frac{\ln \sum \limits_{i=1}^{M(r)} p_{i}^{q}}
{\ln r}, \label{zobecnena dimenze}
\ee
where $p_i=N_i/N$ is the probability that a~point on the attactor falls into
the i-th cube of side $r$, $N$ is the number of the all points on the attractor
and $N_i$ is the number of points in the i-th cube. One obtains the
fractal dimension $d_f$, the information dimension $d_1$ and the
correlation dimension $d_2$ for $q$ tending to $0$, $1$ and $2$, respectively.
Generally $d_f=d_0 \geq d_1 \geq d_2 \dots$, except the case of
uniform distribution of points on the attractor.
\section{Entropy}
The Kolmogorov K-entropy is another important characteristic which describes
a~degree of chaoticity of the system. The entropy gives the average rate of
information loss about a~position of the phase point on the attractor.
It is well-known that
\begin{itemize}
\item $K=0$ in an ordered system
\item $K$ is infinite in a~random system
\item $0<K< \infty$ in a~chaotic (deterministic) system
\end{itemize}
The generalized entropy $K_q$ can be defined like the generalized dimension.
Let ${\bf Y}(t)$, $t>0$ be the trajectory of the dynamical
system in an \hbox{$n$-dimen}sional phase space sampled at discrete time intervals $\Delta t$.
Let us divide the phase space into the $n$-dimensional hypercubes of side $r$.
Let $P_{i_1i_2 \cdots i_N}$ be the joint probability that the trajectory
${\bf Y}(t)$ on the attractor subsequently visits cubes $i_1$, $i_2$, $ \cdots $, $i_N$
at times $t=\Delta t$, $t=2 \Delta t$, $\cdots$, $N \Delta t$. The generalized
entropy is then
\be
K_q=-\lim_{r \to 0} \lim_{\Delta t \to 0} \lim_{N \to \infty}
\frac{1}{N \Delta t} \frac{1}{q-1}
\ln \sum_{i_{1}i_{2} \cdots i_{N}} P_{i_{1}i_{2} \cdots i_{N}}^{q}.
\label{zobecnena entropie}
\ee
One obtains the topological entropy $K_0$, the Kolmogorov entropy $K$
and the $K_2$-entropy for $q$ tending to $0$, $1$ and $2$, respectively.
The $K_2$-entropy is a~lower bound for the Kolmogorov entropy and it is used as its
estimate in most cases because the $K_2$-entropy is easily extracted from an
experimental measurement [17].\par
There are other invariants used for the description of the
chaotic signals: the spectrum of Lyapunov exponents [34], [42], [43],
the Lyapunov dimension [7], [22],
the $\Omega$-dimension [15],
etc.
\section{Estimating from experimental data}
\subsection{Phase space reconstruction}
One of frequent problems in experimental practice is that one has only
a~scalar signal
(generated by the dynamical system) and no governing equations.
The scalars are for example temperature or pressure time series.
Therefore the attractor has to be reconstructed in an artificial phase
space. A~method of time delay coordinates has been
suggested by Takens [37] for this purpose. The main idea of this
procedure is based
on the fact that the phase space variable $x(t)$ contains information about
remaining phase space variables. The $m$-dimensional signal ${\bf Y}(t_i)$
is composed of the scalar series $x(t_i)$ measured at constant sampling
time intervals $\Delta t=t_{i+1}-t_{i}$ as follows
\be
{\bf Y}(t_i) \equiv \left \{ x(t_i), \, x(t_i+ \tau), \, \dots, \, x(t_i+(m-1) \tau) \right \},
\label{rekonstrukce}
\ee
where $\tau$ is an appropriate time delay (which is an integer multiple
of the sampling
time $\Delta t$) and $m$ is an embedding dimension. \par
An important question is then how to choose the embedding dimension and the time delay.
The procedure of finding the embedding dimension $m$ is to
increase
$m$ and to compute the fractal dimension
(or another invariant) for every embedding dimension until the
fractal dimension remains almost constant.
This value of the fractal dimension of the reconstructed attractor is
considered to be equal to the one of the original attractor. Usually
$m \geq 2d_f$ is sufficient [44]. \par
However, the above method is awkward in the case of a~high-dimen\-sional system
and/or in the case of a~large time series. In addition, the
determination of the constant level of the embedding dimension is subjective
and depends on quality of the experimental data. Therefore Broomhead
\& King [3] have suggested another method for the choice of the embedding
dimension based on a~singular value decomposition. Recently Breedon
\& Packard [2] have designed {\it fuzzy} delay coordinate reconstruction for
the data sampled nonuniformly in time. \par
For an infinite amount of noise-free data, the time delay $\tau$ can be
chosen almost arbitrarily. However, when the data are noisy and limited
in number there is uncertainity similar to the one in the choice of
the embedding dimension. The choice of the time delay should guarantee an
independence of the artificial phase space coordinates. The time delay is
usually selected with respect to an autocorrelation function of the data.
Different authors have defined the time delay as the lag at which the
autocorrelation function attains a~certain value. For instance,
\begin{itemize}
\item $1/e$ (Zeng et al. [44])
\item $1/10$ (Tsonis \& Elsner [38])
\item $0$ (Tsonis et al. [41]).
\end{itemize}
Other authors take into account the embedding dimension $m$ and/or 
a~dominant periodicity $T$ of the signal:
\begin{itemize}
\item $T/m$ (Tsonis et al. [41])
\item $w/(m-1)$, where $w$ is the first local minimum of the autocorrelation
function or the first lag for which the autocorrelation function passes
through zero (Poveda \& Puente [29]).
\end{itemize}
\par
However, the autocorrelation function measures only a~linear dependence.
Therefore Fraser \& Swinney [14] have suggested to choose the time delay
by so called {\it mutual information} which measures the general
dependence. But this method is not yet widely used.
\subsection{Estimating the dimension}
The information and the correlation dimensions are the most commonly computed
invariants. Given the $m$-dimensional signal (\ref{rekonstrukce}), one defines
a~local correlation integral $C^m(i,r)$ as follows
\be
C^{m}(i,r)=\frac{1}{M-1} \sum_{\begin{array}{c}
                 j=1 \\ i~\neq j
                               \end{array}}^{M} \theta (r-\| {\bf Y}_i- {\bf Y}_j \|),
\label{lokkorint}
\ee
where $M=N-(m-1) \tau$ and $N$ is the number of data points in the original
one-dimensional signal.
$\theta(x)$ is a~Heaviside step function defined by
\be
\theta(x)=\left \{ \begin{array}{ll}
                    1 & \mbox{for $x \geq 0$} \\
                    0 & \mbox{for $x < 0.$}
                   \end{array}
          \right.
\label{heavfukce}
\ee
$\| \dots \|$ denotes an appropriate norm in the phase space, usually
the Euclidean one.
\be
d_1= \lim_{r \to 0} \lim_{M \to \infty} \frac{ \ln C^{m}(i,r)}{ \ln r}.
\label{infordim}
\ee
Experimentally, $d_1$ may be obtained as the slope of the linear part of
the curve $\ln C^m$ versus $\ln r$. The result should be independent
of $i$ for a~sufficiently large $M$. Eckman \& Ruelle [8] have discussed
the cases when this approach may fail. \par
To improve the statistics one may average $d_1(i)$ over several $i$.
Grassberger \& Procaccia [16] have used averaging over all $i$
and they have obtained the correlation dimension
\be
d_2= \lim_{r \to 0} \lim_{M \to \infty} \frac{ \ln C^{m}(r)}{ \ln r},
\label{koreldim}
\ee
where $C^{m}(r)$ is the correlation integral defined as folows
\be
C^{m}(r)=\frac{1}{M(M-1)} \sum_{\begin{array}{c}
                                    i,j=1 \\ i~\neq j
                                \end{array}}
                                    ^{M} \theta (r-\| {\bf Y}_i- {\bf Y}_j \|).
\label{korint}
\ee
\par
In practice $d_2$ is obtained by plotting
$\ln C^m (r)$ versus $\ln r$ and determining the slope of the curve
between $r_{min}$ and $r_{max}$.
For $r$ less than $r_{min}$ there is poor statistics due to sparseness
of points and for $r$ greater than $r_{max}$  nonlinear effects deviate
the dependence from the straight line.
Only the interval $(r_{min},r_{max})$ is ``meaningful''. No theoretical
prescription exists for selecting the bounds $r_{min}$ and $r_{max}$. It is
left to researcher's judgement. An inappropriate selection may substantially
devalue the result. The meaningful scaling region may be masked if the number
of data points $N$ is not sufficiently large and/or if the embedding
dimension $m$ exceeds a~critical embedding dimension $m_c$ [9], [39].
We will show below that a~faulty conclusion may
be obtained if the time series is too small, although
the scaling region is evident.
An increase of the number of points in the time series by means of
interpolation does not help and produces a~spurious slope at small $\ln r$
[33]. \par
The question of how many points are enough is a~widely discussed problem
[9], [10], [12], [24], [33], [35], [41], [44].
Many formulae determining a~minimal number of data points $N_{min}$ for
a~reliable estimate have been established. However, it has been recently
demonstrated that some of these formulae are much strict or erroneous
[33], [35]. Nerenberg \& Essex [24] have concluded that
\be
N_{min}=\frac{\sqrt{2} \sqrt{\Gamma \left( m/2 +1 \right) }}
         {\left( A~\ln k~\right)^{( m+2 ) /2}}
     \left \{ \frac{2 \left( k-1 \right) \Gamma \left[ \left( m+4 \right)
         /2 \right] }
         {\left[ \Gamma \left( 1/2 \right) \right] ^2
         \Gamma \left[ \left( m+3 \right) /2 \right] } \right \}
         \frac{m+2}{2},
\label{esnermin}
\ee
where $\Gamma(x)$ is the gamma function, $k \equiv r_{max}/ r_{min}$ and
$A$ is permitted error of the estimation.
This formula has been derived under preconception that $m < d_2$.
Tsonis et al. [41] have shown that a~satisfactory estimate of the
correlation dimension of the H\' enon map can be obtained in the embedding
dimension $m=8$ using 2000 data points only. It is much smaller number
of points
than the formula (\ref{esnermin}) indicates. Therefore the need for
data has not to be the same for $m>d_2$ as for $m<d_2$. The minimal number
of data points $N_{min}$ would depend on the type of the attractor
[41]. \par
It was mentioned above that $d_1 \geq d_2$ and that the information
dimension is the lower bound for the fractal dimension. The inequality
occurs in the limit case $M \to \infty$ in the formulae (\ref{infordim})
and (\ref{koreldim}) but it is reasonable to assume that the equality
holds good for the data limited in number [8].
Let us therefore consider them to be mutually equal and refer the common
value as the fractal dimension $d_f$ henceforth.
\subsection{Estimating the entropy}
Let us pay attention to estimating the entropy. Grassberger \& Procaccia
[17] have proposed that
\be
K_2 \sim \lim_{m \to \infty} \lim_{r \to 0} K_{2}^{m}(r),
\label{k2entrop}
\ee
where
\be
K_{2}^{m}=\frac{1}{k \Delta t} \ln
         \frac{C^m (r)}{C^{m+k}(r)}
\label{k2mentrop}
\ee
and $k$ is a~sufficiently small integer number. In practice one can not
satisfy the limits in the above formulae. Therefore the saturation value
of $K_2^m$ as $m$ increases is regarded as $K_2$. An extrapolation
formula for $m \to \infty$ has been used in the original paper of
Grassberger \& Procaccia [17] but its explicit form has not been
mentioned. The use of a~maximum norm instead of the Euclidean norm
in the equations (\ref{lokkorint}) and (\ref{korint}) leads to improvement
of the $K_2^m$ convergence but an anisotropy of the maximum norm may cause
an underestimation of the correlation dimension [13]. \par
There is a~certain degree of uncertainty in estimating the dimension and the
entropy from experimental time series. The algorithms are usually
reliable when dealing with artificial time series generated by means of
a~computer. However, the results need a~very careful interpretation
when dealing with the real measured data containing
inherent noise and limited in number. \par
Moreover, the determination of a~finite noninteger value of the fractal
dimension and/or a~finite value of the $K_2$-entropy greater than zero
is not sufficient to claim that the system has a~strange attractor.
Osborne et al. [26], Osborne \& Provenzale [27],
Provenzale et al. [30] and Provenzale et al. [31] have presented a~class
of random fractal (coloured) noise with the finite noninteger
correlation dimension and the converging $K_2$-entropy estimates.
Such noise is characterized by a~power-law power spectrum
\be
P(\omega_k) \sim \omega_k^{-\alpha}, \qquad 1< \alpha <3
\ee
and its correlation dimension is
\be
d_2=\frac{2}{\alpha -1}.
\ee
Therefore necessity to distinguish between this class of noise and
deterministic dynamics arises
in the analysis of experimental data. Some methods have been suggested
by Provenzale et al. [31], Pavlos et al. [28] and Tsonis \& Elsner
[40]. The tests are based on generating appropriate surrogate series
and comparing their characteristics with the ones of the original
series.
\section{Strange attractor in the atmosphere}
Application of the deterministic chaos theory to atmospheric phenomena
has been motivated by the attempt to reveal their possible low-dimensional
nature and reduce the number of variables needed for describing the
phenonema. Indeed, the first studies referred to low-dimensional
($d_f$ between $3$ and $8$) weather and climate attractors
[9], [19], [23], [25], [32], [38].
Later these results have been criticized by
other authors [33], [44]. They have argued by
considerable complexity of the atmosphere. Therefore they have not believed
in the low-dimensional character of the atmosphere. The critics have
considered the low estimates of the dimension as a~consequence of
the limited number of data points. Recently published papers have not thrown
light on the problem. They have reported upon the low-dimensionality
[1] on the one hand and upon the high-dimensionality
[12] of the atmosphere on the other hand. \par
Lorenz [21] has presented an imaginative explanation. He has shown
that a~meaningful estimate is possible by using only a~relatively small
number of data points if a~variable strongly coupled with the rest of the
variables of the system is used. However, he has shown that the fractal
dimension estimate is undervalued if a~weakly coupled variable is used.
In this case rather the dimension of a~subsystem is measured. \par
Note that the existence of an attractor is presupposed a~priori
in the methods described above. This preconception is reasonable in the
case of the atmosphere because it is hard to imagine that the weather is
completely governed by some kind of randomness.
\subsection{Used data and methods}
The data utilized in our experiment include mean daily temperature of air
over a~period of 215 years (1 January 1775 - 31 December 1989) recorded
at the Klementinum Observatory, Prague, Czech Republic.
The total number of data points is 78527 which is the largest
sample measured at one meteorological station, used for such an analysis.
The reader may have doubts about the quality of the data, especially
at the beginning of the observation. However, Hlav\' a\v c [18] has paid
extensive attention to homogenization of the series, especially with
the respect to accuracy of the measurement in the 18th and the 19th centuries.
\par
In addition to the entire temperature time series, different modifications
of it
are explored. They are cut versions, a~series without annual course
and its cut versions, a~smoothed series, a~time-differenced series and
``seasonally adjusted'' series. Together it gives 38 versions of the original
series. \par
Because of the fact that the sample is very extensive the correlation
integral (\ref{korint}) is replaced by
\be
C^{m}(r)=\frac{1}{I(M-1)} \sum_{i=1}^{I} \sum_{\begin{array}{c}
                                    j=1 \\ i~\neq j
                                \end{array}}
                                    ^{M} \theta (r-\| {\bf Y}_i- {\bf Y}_j \|),
                                 \qquad I=250
\label{mujkorint}
\ee
in order to reduce a~heavy computational burden. The selection of $i$ is
led by the idea of roughly uniform distribution of indices $i$
between $1$ and $M$:
\be
i=\left[ \frac{M}{I+1} \right] + p \left[ \frac{M}{I} \right],
                                 \qquad p=0,1, \dots ,I-1,
\ee
where $[x]$ denotes the integer part of $x$. \par
The fractal dimension $d_f$ is estimated from the straight line slope
fitted to the ``meaningful'' range of the plot $\ln C^m(r)$ versus $\ln r$.
The fitting is carried out by the least-squares regression
in the interval \hbox{$< \ln r_1, \ln r_n>$}, \hbox{$\ln r_{min} \leq \ln r_1 < \ln r_n \leq \ln r_{max}$}.
In order to objectify the selection of the appropriate
scaling region the bounds $\ln r_1$ and $\ln r_n$ are determinated
so that the term
\be
\left| \frac{ \frac{1}{n} \sum \limits_{k=1}^{n} \left( \ln r_k -\overline{\ln r} \right)
                    \left(\ln C_k^m - \overline{\ln C_k^m} \right)}
     {\sqrt{\left[ \frac{1}{n}
      \sum \limits_{k=1}^{n} \left( \ln r_k -\overline{\ln r} \right) \right]
      \left[ \frac{1}{n} \sum \limits_{k=1}^{n} \left(\ln C_k^m - \overline{\ln C_k^m} \right)
      \right]}} \right|
      \equiv \left| r^{m}(r_1, r_n) \right|,
\label{korkoef}
\ee
where
\be
\overline{\ln r}=\frac{1}{n} \sum_{k=1}^{n} \ln r_k,  \qquad
\overline{\ln C_k^m}=\frac{1}{n} \sum_{k=1}^{n} \ln C_{k}^{m},
\ee
is maximal. This means that the estimate is done in that coherent
part of the plot in which the absolute value of the linear regression
\hbox{coefficient}
$r^m(r_1,r_n)$ is maximal in every embedding dimension. The restriction
\hbox{$n \geq 15$} is introduced for the scaling region in order to be sufficiently
large. \par
The aforementioned method was tested by H\' enon map and gave very promising
results. We obtained $d_f=1.21$ for $m=10$, $N=5000$. Results were more
precise for lower embedding dimensions ($d_f=1.25$ for $m=4$, $N=5000$).
\subsection{Estimating the dimension of the climate attractor}
The fractal dimension estimates are performed for the following series:
the mean daily temperature series of air recorded from 1 January, 1775 until
31 December, 1989 (hereafter the original series), fifteen
consecutive 15-year-long versions of the original series
(hereafter the cut series of the original one), the original series without
annual course (hereafter the filtered series), fifteen
consecutive 15-year-long versions of the filtered series
(hereafter the cut series of the filtered one), a~spring series,
a~summer series, an autumn series, a~winter series, the original series
smoothed by five-day moving averages (hereafter the smoothed
series), a~first time-difference of the original series (hereafter
differenced series). \par
In Figure 1 one may see the course of the autocorrelation functions for the
individual series. Table 1 shows the lag at which the autocorrelation
functions for the first time attain the values $1/e$, $1/10$ and $0$.
The threshold values of $1/e$ as the time delay $\tau$ is used in this study.
\subsubsection{Original series}
Figures 2a and 2b present the $\ln C$-$\ln r$ diagram for different
values of the embedding dimension. One can see two linear parts for
the embedding dimension greater than 4. There is the scaling region (1)
at an intermediate part of $\ln r$. The slope of this region increases
as the embedding dimension increases and for $m=50$ the slope attains
the value of $17.5 \pm 0.3$. The error represents the 90-\% confidence
interval of the linear least-squares fit. The second linear part (denoted as (2)) is evident
for a~larger $\ln r$ earlier than the nonlinear effects cause the deviation
from constancy. The scaling region (2) has a~very gradual
slope with a~fast convergence to the value $1.6$ with the error less than $0.005$.
The saturation is already reached for the embedding dimension $m=4$,
unlike the scaling region (1) when the ``saturation'' is reached between
$m=46$ and $50$ (see Figures 3a and 3b). The quotation marks denote that
we are not sure whether further increasing of the embedding dimension
would be accompanied  by increasing of the slope! \par
In spite of the fact that the range of the smallest $\ln r$
is biased by fluctuations,
an additional region appears there for the embedding dimension greater
than $20$. This ``scaling region'' is divided into two parts (3) and (4)
for the embedding dimension greater than $30$. The slope of the lower
region (3) decreases and the slope of the upper region increases as
the embedding dimension increases. The slopes of the scaling regions
(3) and (4) are $16.1 \pm 0.8$ and $2.7 \pm 0.1$, respectively,
for the embedding dimension $m=50$. These scaling regions do not saturate
their slope for $m \leq 50$ and they are not so clearly expressed as the
scaling regions (1) and (2). For this moment let us make do with
identification of the two clearly linear (see correlation coefficient in
Table 2)
scaling regions (1) and (2) with the slopes $17.5$ and $1.6$, respectively.
\subsubsection{Filtered series}
We could suspect that the scaling region (2) with the slope $1.6$ is
related to seasonal variations of temperature. In order to minimize the effect
of seasonal changes of temperature a~mean temperature is computed
for each day over the original series and these daily means are
subtracted from each daily value of temperature. For instance, the mean
temperature $ \bar{T}_{15.6.}^{1775-1989}$ of 15 June is computed
by temperature averaging over the all 15 Junes and then this mean is
subtracted from temperature of each 15 June. The filtered series is
obtained in this way. \par
One can see the linear part (1) in the $\ln C$-$\ln r$ diagram in
Figure 4 with a~fast slope. This scaling region probably corresponds
to the scaling region (1) of the original series. The slope ``converges'' to
the value of $21.4 \pm 0.4$ for the embedding dimension $m$ between
$46$ and $50$ (see Figure 5). There are also the scaling regions (3) and
(4) for smaller $\ln r$. They arise for $m>30$, that means for
the higher value of the embedding dimension than in the case of the
original series. The slopes of these scaling regions do not saturate
for the embedding $m \leq 50$. The slope of the scaling region (4)
decreases (for $m=50$ is equal to 4.1) and the slope of scaling region (3)
increases (for $m=50$ is equal to 13.9)
as the embedding dimension increases. \par
The scaling region with the gradual slope in the range of larger $\ln r$ is
completely missing. This supports our hypothesis that the scaling region (2)
in the correlation integrals
of the original series is consequence of the seasonal course of temperature.
\subsubsection{Cut series}
Both the original and the filtered time series are divided into the fifteen
15-year-long consecutive intervals. The length of any cut series is
$5478$ or $5479$ data points. \par
Dependence $\ln C^m$ on $\ln r$ is qualitatively the same for
all thirty cut series as well as for the series from which they are created.
The fast convergence of the slope to the value of $1.6$ is registered
in the region of larger $\ln r$ for any cut series of the original one.
The slope of the scaling region which was denoted as (1) in the case of
the original and filtered series converges to the values between
$8.5$ and $10.4$ for the embedding dimension between
$22$ and $38$ according to the selection of 15-year interval.
This is considerable
underestimation of the slope in comparision with the original series. \par
The same undervaluation can be observed for any cut series of the filtered
one. The convergence of the slope of the scaling region (1) to the values
between $14.3$ and $16.2$ is reached for
the embedding dimension between $34$ and $38$
according to the selection of 15-year interval. The series of the
years 1835-1849, 1865-1879, 1880-1894, 1895-1909 and 1975-1989 are
exceptions because no clearly saturated values of the slopes are
reached and the slopes are greater than $17.1$ for the embedding
dimension $m=50$. \par
Therefore we conclude that a~15-year-long series of the mean daily temperature
of air is not adequate to estimate a~high fractal dimension based on
the Grassberger \& Procaccia algorithm. This is strongly marked for the
original series. In this case the obtained values of the fractal dimension
(between $8.5$ and $10.4$) are very close to estimates of other authors
[19], [38] who have used shorter
series than ours. However, one has to keep in mind that their series have
not been always the temperature ones (see the aforementioned Lorenz's
experiment [21]). \par
We have generated a~time series by means of gaussian random process
in order to exclude an artificial saturation of the slope of the scaling
regions (1) due to insufficient number of the data
in the case of the full time series. The length of this tentative series
was $78527$ data points, the mean was zero and the variance was $3.7$.
These parameters are close to the ones of the filtered series. For the
random series no saturation has been observed even if the calculation
has been carried out to the embedding dimension $m=80$.
\subsubsection{Seasonal series}
Seasonally adjusted series are created from the original one. They are
the spring one (1 March - 31 May), the summer one (1 June - 31 August),
the autumn one (1 September - 30 November) and the winter one (1 December
- 29 February). \par
The linear parts (1) are clearly evident in the $\ln C^m$-$\ln r$ diagrams
for all seasonal series (see Figure 6). The slopes of these scaling
regions are again large and converge as the embedding dimension increases.
The ``saturation'' values are $16.4 \pm 0.3$? for the spring series,
$20.3 \pm 0.3$? for the summer series, $15.9 \pm 0.3$ for the autumn series
and $17.7 \pm 0.3$? for the winter series. The question marks after the
values indicate uncertainty of the saturation. \par
The scaling regions (2) appear above the scaling regions (1) only in
the case of the spring and autumn series. Their slopes converge towards
the value of $4.5 \pm 0.1$ for the spring series and towards the value
of $3.8$ with the error less than $0.01$  for the autumn series. No similar
scaling regions are detectable in the summer and winter series. \par
The existence of the scaling regions (2) in spring and autumn
supports the above mentioned hyphothesis that the gradual slope in the
correlation integrals in the range of large $\ln r$ is created by seasonal
variability of weather. Moreover, it is not obviously simple annual
course because of their absence in the case of summer and winter.
\subsubsection{Differenced series}
For a~system governed by low-dimensional dynamics, the fractal dimension
is the same for the original signal as well as for its first difference.
However, the first difference of stochastic signal has often much larger
value of the fractal dimension than the original signal
[31].
The first time-differenced series is created from the
original one in order to exclude or confirm the deterministic character
of the data. \par
From Figure 8a one can see one and only scaling region in the correlation
integrals. This is quite different dependence than in the earlier cases.
The slopes of the scaling region oscillate at first between $19.4$ and
$20.3$ for the embedding dimension between $34$ and $ 42$, and later
between $20.5$ and $26$ for the embedding dimension between $50$ and $60$
(see Figure 8b). The value of the fractal dimension acquired from the
differenced series is closer to the one obtained from
the filtered series than
from the original series (see Table 2). \par
One can immediately exclude coloured noise with high confidence
owing to the course of the autocorrelation functions (see Figure 1a and 1c).
The autocorrelation function of the coloured noise series decreases very
slowly and in the case of unlimited number of data points never reaches
the value of zero [41]. The autocorrelation function
of the differenced series looks like the one of white-noise. \par
Although random fluctuations are present in the temperature series
we suppose that they are not responsible for the origin
of the scaling regions (1)
from which the fractal dimension values of $17.5$ and $21.4$ are
estimated. We conclude that from the above mentioned experiment with gaussian
random process. We presume that one meets high-dimensional dynamics and
that the oscillations of the correlation integral slopes between $19.4$
and $20.3$ and then around the value of $22.5$ are caused by
presence of inherent noise. It is well-known ([1], [28], [31])
that differencing acts as
an amplifier of noisy components
but weakens low-frequency components included in the original signal.
In this
way we explain the absence of another scaling region corresponding to
a~low value of the fractal dimension in the case of differenced series.
Pavlos et al. [28] have observed similar low-dimensional chaotic
component cut off due to differencing.
\subsubsection{Smoothed series}
The high-frequency oscillations can be removed by means of averaging of the
original signal over an appropriate time interval. In this study 5-day
moving averages are applied. \par
The correlation integrals computed from the series smoothed in this way are
shown in Figures 9a and 9b. The dependence is similar to the one for the
original series (see Figures 2 and 9). However, the individual scaling
regions are more evident than for the original series. This is particularly
valid for the scaling regions (3) and (4). This phenomenon is easy to
understand because random fluctuations reside in the range of smaller
scales and the smoothing removes them. \par
The values of the fractal dimension estimated from the scaling regions (1)
and (2) are slightly smaller than the values from the non-smoothed series.
We get the values $16.2 \pm 0.3$ and $1.5$ with the error less than
$0.004$ from the scaling regions (1) and (2), respectively. The decrease
of the value of the fractal dimension for smoothed series has
been observed by Pavlos et al. [28] as well. We assume that the decrease
is due to high-frequency random noise. If an $m$-dimensional signal is
reconstructed from one-dimensional random sequence then the cloud of
\hbox{$m$-dimensional} points fills out completely the $m$-dimensional phase space.
If the signal is composed of the deterministic and stochastic
components then the stochastic one preserves its tendency to fill out
the phase space. This causes larger smearing of the trajectories in the phase
space and one then measures higher dimension than in the purely deterministic
case.  \par
Let us pay attention to the scaling regions (3) and (4). The slopes of the
scaling regions (3) and (4) oscillate between $14.8$ and $17.7$,
and between $1.5$ and $1.7$, respectively. These values are very close
to the ones
of the scaling regions (1) and (2) (see Figures 10a and 10b). That looks like
the dynamics is duplicated in the smaller scales. Could it be possible
that the
scaling regions (3) and (4) are manifestation of additional subsystems which
reside in very small scales and, therefore, they are not clearly detectable
in the signal contaminated by high-frequency noise? However, it is quite
possible that used algorithm introduces this ``symmetry'' artificially
if it is applied on the extreme embedding dimension. Many experiments
have to be carried out in order to confirm one of these hypotheses or
another different one.
\subsection{Estimating the entropy
of the climate attractor}
The $K_2$-entropy is estimated for the original, filtered, smoothed
and all seasonally adjusted series. The formula (\ref{mujkorint}) is
again used instead of the correlation integral (\ref{korint}). In order to
reduce fluctuations and improve the statistics the formula (\ref{k2mentrop})
is averaged over five values computed for different embedding
dimensions. This yields
\be
K_2^m (r) \equiv \frac{1}{L} \sum_{l=1}^{L} \frac{1}{2 l \Delta t}
      \ln \frac{C^{m-2l}(r)}{C^{m}(r)}, \qquad L=5.
\label{k2l5}
\ee
Dependence of $K^m_2$ on the embedding dimension $m$ is approximated
by means of least-squares fit by the function
\be
f(x)=a+ \frac{b}{x^c}.
\ee
This function converges to $a$ for $c>0$ and $x \to \infty$.
Therefore
\be
K_2^m (r)=K_2(r)+\frac{b}{m^c}, \qquad c>0.
\label{extra1}
\ee
$b$ a~$c$ are real parameters, $m$ is the embedding dimension and $K_2(r)$
is the entropy which depends on the selection of the scaling region.
The formula (\ref{extra1}) describes the dependence of $K^m_2$
on $m$ very appropriately (see Figures 11, 12, 13 and Table 3). \par
The value of an error-doubling time is more illustrative of the predictability
of the atmosphere than the value of the $K_2$-entropy. The error-doubling
time is defined as follows
\be
T_2= \frac{\ln 2}{K_2}.
\label{T2}
\ee
\par
The estimates given in Table 3 are obtained from the scaling regions (1)
and (2) of the correlation integrals. It has not been possible to acquire
reliable estimates from the scaling regions (3) and (4) owing to large
oscillations of $K_2^m$ in the range of the smallest $\ln r$. \par
Let us introduce results of the estimates from the scaling regions (1) first.
The error-doubling times of the systems reconstructed from
the original and filtered series are $4.7 \pm 0.9$ days
and $5.9 \pm 0.9$ days, respectively. These values, especially the
latter one, are very
close to the time scale of the synoptic fluctuations. \par
The transition-seasons (spring and autumn)  exhibit increase of the
value of the $K_2$-entropy and decrease of the value of the error-doubling
time ($T_2=4.0 \pm 0.4$ days for spring and $T_2=3.6 \pm 0.4$ for autumn).
The greater values of the error-doubling time are obtained for the
summer and winter seasons ($4.7 \pm 0.4$ days for summer and $4.7 \pm 0.6$
days for winter). These values are the same as for the entire year.
Therefore the most difficult prediction should be expected in spring and
autumn. This should not be surprising if one remembers
changeable weather during
these transition-seasons which is connected with the change of general
circulation from summer to winter and vice versa. The estimate from
the smoothed series yields the value of $T_2=6.1 \pm 0.3$ days. \par
As mentioned above, the scaling regions (2) with the gradual
slope are revealed in the ranges of larger $\ln r$ in the original,
smoothed, spring and autumn series. The results of the estimates of
the $K_2$-entropy from these regions are shown in Figures 14 and 15
and in Table 3. The values for spring are not involved because
a~``plateau''
in the $K_2^m$-$\ln r$ diagram is clearly  detectable only for the
embedding dimensions greater than $52$ and the extrapolation by means of
the formula (\ref{extra1}) is unreliable. The error-doubling times
computed from the scaling regions (2) are surprisingly high even if
the values of $K_2^m$ for $m=50$ are not so high ($28.1$ days for the entire
year, $27.0$ for the smoothed series, $15.5$ for autumn).
However, the formula
(\ref{extra1}) gives the relatively extreme estimates which need further
verification. \par
For this purpose, it is possible to use the maximum norm
instead of the Euclidean
one. Improvement of the $K_2$-entropy calculation can be reached by means of
dimension scaled distances technique [13]. In this way,
one could avoid the extrapolation. Further possibility is to
utilize another extrapolation formula instead of (\ref{extra1}).
Formula
\be
K_2^m=K_2 + \frac{d_2}{2 \Delta t} \ln
     \frac{m+1}{m},
\label{frank}
\ee
which has been proposed by Frank et al. [13] should be suitable. However,
the formula needs some modification in our case because the averaging
is introduced by means of (\ref{k2l5}). Therefore one has
\be
K_2^m=K_2+ \frac{d_2}{4L \Delta t}
      \sum_{l=1}^{L} \frac{1}{l}\ln \left( \frac{m+2l}{m} \right) , \qquad L=5
\label{raidlfrank}
\ee
instead of (\ref{frank}).
\par
Some of the aforementioned improvements of the $K_2$-entropy calculation
will be carried out in future.
\section{Discussion and conclusions}
Some ideas from nonlinear time series analysis have been used to study
the dimensionality and the predictability of weather system.
The analysis of the
series of mean daily temperature of air and the series generated by it
has been carried out. The fractal dimension and the $K_2$-entropy estimates
have been performed by means of modified algorithms of Grassberger \&
Procaccia. Since a~careless selection of the scaling region
in the $\ln C^m$-$\ln r$ diagram may often lead to erroneous conclusions
the attempt to objectify the choice has been introduced by means of
maximization of the formula (\ref{korkoef}) in every embedding dimension.
The formula (\ref{extra1}) has been utilized for the extrapolation
of $K_2^m$ for
$m \to \infty$ and formula (\ref{raidlfrank}) has been devised for the same
aim. \par
The series has exhibited multiscaled character. Several scaling regions of
the correlation integrals have been revealed. The first one
possessed the gradual slope in the range of larger $\ln r$. The second one
possessed the fast slope in the range of smaller $\ln r$.
This arrangement can be explained according to Eckman \& Ruelle [8]
as a~product of two dynamical subsystems $A$ and $B$. Suppose that the
subsystems $A$ and $B$ are noninteracting or that the subsystem $B$
evolves independently of $A$, but the evolution of $A$ may depend on $B$.
Let $X_A$ and $X_B$ be the signals of the subsystems
$A$ and $B$, respectively. Suppose that the amplitude $r_A$ of the first signal $X_A$
is less than that of the signal $X_B$. Then one has an information about
the complete system $(A+B)$ in the region $\ln r< \ln r_A$. In the region
$\ln r \gg \ln r_A$ one has an information about the subsystem $B$ only.
The ``knee'' in the correlation integral may also result
from a~purely stochastic
process [31] but this is not our case. Other two
``scaling'' regions have been found in the range of smallest $\ln r$. They
have been most visible for the smoothed series.
Our preliminary results indicate that this regions are due to the temporal
correlations between nearby points of the series. The detailed calculation
will be published later (see also [45]).
\par
We conclude that the temperature series has been formed as a~result of
activity of at least two subsystems. One subsystem has the fractal
dimension greater
than $17$, the other one is low-dimensional ($d_f=1.6$). The error-doubling
time of the high-dimensional subsystem is comparable with the time
scale of the synoptic fluctuations (from four to six days). The
low-dimensional subsystem exhibits much larger potential predictability.
In spite of some problems in the $K_2$-entropy estimation and
the surprisingly high
obtained values of the error-doubling time,
we suggest that the value of $T_2$ for the low-dimensional subsystem is
of the order of a~few weeks. The low-dimensional subsystem is probably
related
to the seasonal variability of weather because it was not detectable
in the series without annual course and in some of seasonally adjusted
series. \par
The values of the $K_2$-entropy of the high-dimensional subsystem demonstrate
lower predictability time in  transition-seasons (spring and autumn). \par
We conclude that 15-year-long mean daily temperature series are not
sufficient for the reliable estimation of a~high fractal dimension
based on the Grassberger \& Procaccia algorithm. The values of the fractal
dimension of the high-dimensional subsystem estimated from the cut series
have been reduced roughly by half in comparison with the full time series.
\par
The temperature series has not purely deterministic character. This has been
demonstrated by decreasing of the fractal dimension and the $K_2$-entropy
for the smoothed series and also by increasing of them for the differenced
series,
in comparison with the original one. Further
tests should be carried out in order to distinguish precisely between the
deterministic component and noise. \par
The estimates of the dimension and the entropy are only some of many
steps which ought to lead to more complete understanding of the
dynamics of the atmospheric processes. Our attempt should continue by
calculation of other invariants. It is desirable to carry out the estimates
from series of other meteorological elements. Observations from different
meteorological stations should be used in order to remove local impacts.
The research ought to culminate in attempt at nonlinear prediction
[5], [11], [36], [40]. Here an interesting question arises in relation
to the low-dimensional subsystem with higher potential predictability.
Namely, is a~successful long-term prediction possible if the high-dimensional
subsystem is excluded from the signal? Moreover, it is not clear whether
(and how much) such prediction would be different from a~climatological
normal.
\newline
\newline
{\it Acknowledgements}. Author would like to thank Dr.\,J.\,Hor\' ak from
the Institute of Atmospheric Physics AS CR for his support and helpful
discussions
and P.\,Habala from University of Alberta, Canada for his comments on
language.
I~am also grateful to P.\,Koln\'\i{}k and M.\,Dov\v ciak for their
help in preparation of the manuscript. Computations were carried out on
the Institute of Physics AS CR Cray computer.
\newpage
\section*{References}
\begin{description}
\item{[1] Bountis T., L.Karakatsanis, G.Papaioannou, G.Pavlos,
     Ann.Geophys., {\bf10}, (1993), 947}
\item{[2] Breedon J.L., N.H.Packard,
     Physica D, {\bf 58}, (1992), 273}
\item{[3] Broomhead D.S., G.P.King,
     Physica D, {\bf 20}, (1986), 217}
\item{[4] Bunde A., S.Havlin,
     Fractals and disordered systems, Springer-Verlag, Berlin,
     Heidelberg, 1991}
\item{[5] Casdagli M.,
     Physica D, {\bf 35}, (1989), 335}
\item{[6] Dvo\v r\' ak I., J.\v Si\v ska,
     Pok.mat.fyz.astro., {\bf 36}, (1991), 73}
\item{[7] Dymnikov V.P., A.N.Filatov,
     Stability of large-scale atmospheric processes, Gidrometeoizdat,
     Leningrad, 1990}
\item{[8] Eckman J.-P., D.Ruelle,
     Rev.Mod.Phys., {\bf 57}, (1985), 617}
\item{[9] Essex C., T.Lookman, M.A.H. Nerenberg,
     Nature, {\bf 326}, (1987), 64}
\item{[10] Essex C., M.A.H. Nerenberg,
     Proc.R.Soc.Lond. A, {\bf 435}, (1991), 287}
\item{[11] Farmer J.D., J.J.Sidorowich,
     Phys.Rev.Lett., {\bf 59}, (1987), 845}
\item{[12] Fraedrich K., R.Wang,
     Physica D, {\bf 65}, (1993), 373}
\item{[13] Frank M., H.R.Blank, J.Heindl, M.Kaltenh\" auser, H.K\" ochner,
     W.Kreische, N.M\" uller, S.Poscher, R.Sporer, T.Wagner,
     Physica D, {\bf 65}, (1993), 359}
\item{[14] Fraser A.M., H.L.Swinney,
     Phys.Rev. A, {\bf 33}, (1986), 1134}
\item{[15] Gaponov-Grekhov A.V., M.I.Rabinovich, I.M.Starobinets,
     M.Sh.Tsimring, V.V.Chugurin,
     Chaos, {\bf 4}, (1994), 55}
\item{[16] Grassberger P., I.Procaccia,
     Phys.Rev.Lett., {\bf 50}, (1983), 346}
\item{[17] Grassberger P., I.Procaccia,
     Phys.Rev. A, {\bf 28}, (1983), 2591}
\item{[18] Hlav\' a\v c V.,
     Tepeln\' e pom\v ery hl. m\v esta Prahy, \v Cs.stat., Praha, 1937 (in Czech)}
\item{[19] Keppenne C.L., C.Nicolis,
     J.Atmos.Sci., {\bf 46}, (1989), 2356}
\item{[20] Lorenz E.N.,
     J.Atmos.Sci., {\bf 20}, (1963), 130}
\item{[21] Lorenz E.N.,
     Nature, {\bf 353}, (1991), 241}
\item{[22] Marek M., I.Schreiber,
     Chaotic behaviour of deterministic dissipative systems, Academia,
     Praha, 1991}
\item{[23] Mohan K., J.S.Rao, R.Ramaswamy,
     J.Climate, {\bf 2}, (1989), 1047}
\item{[24] Nerenberg M.A.H., C.Essex,
     Phys.Rev. A, {\bf 42}, (1990), 7065}
\item{[25] Nicolis C., G.Nicolis,
     Nature, {\bf 311}, (1984), 529}
\item{[26] Osborne A.R., A.D.Kirwan, A.Provenzale, L.Bergamasco,
     Physica D, {\bf 23}, (1986), 75}
\item{[27] Osborne A.R., A.Provenzale,
     Physica D, {\bf 35}, (1989), 357}
\item{[28] Pavlos G.P., G.A.Kyriakou, A.G.Rigas, P.I.Liatsis, P.C.Trochoutsos,
     A.A. Tsonis,
     Ann.Geophys., {\bf 10}, (1992), 309}
\item{[29] Poveda J., C.E.Puente,
     Bound.Lay.Met., {\bf 64}, (1993), 175}
\item{[30] Provenzale A., A.R.Osborne, R.Soj,
     Physica D, {\bf 47}, (1991), 361}
\item{[31] Provenzale A., L.A.Smith, R.Vio, G.Murante,
     Physica D, {\bf 58}, (1992), 31}
\item{[32] Romanelli L., M.A.Figliola, F.A.Hirsch,
     J.Stat.Phys., {\bf 53}, (1988), 991}
\item{[33] Ruelle D.,
     Proc.R.Soc.Lond. A, {\bf 427}, (1990), 241}
\item{[34] Sano M., Y.Sawada,
     Phys.Rev.Lett., {\bf 55}, (1985), 1082}
\item{[35] Smith L.A.,
     Phys.Lett. A, {\bf 133}, (1988), 283}
\item{[36] Sugihara G., R.M.May,
     Nature, {\bf 344}, (1990), 734}
\item{[37] Takens F.,
     Detecting strange attractors in turbulence, in Lecture notes
     in mathematics, Springer, Berlin, 1981}
\item{[38] Tsonis A.A., J.B.Elsner,
     Nature, {\bf 333}, (1988), 545}
\item{[39] Tsonis A.A., J.B.Elsner,
     J.Climate, {\bf 3}, (1990), 1502}
\item{[40] Tsonis A.A., J.B.Elsner,
     Nature, {\bf 358}, (1992), 217}
\item{[41] Tsonis A.A., J.B.Elsner, K.P.Georgakakos,
     J.Atmos.Sci., {\bf 50}, (1993), 2549}
\item{[42] Wolf A., J.B.Swift, H.L.Swinney, J.Vasto,
     Physica D, {\bf 16}, (1985), 285}
\item{[43] Zeng X., R.Eykholt, R.A.Pielke,
     Phys.Rev.Lett., {\bf 66}, (1991), 3229}
\item{[44] Zeng X., R.A.Pielke, R.Eykholt,
     J.Atmos.Sci., {\bf 49}, (1992), 649}
\samepage
\item{[45] Theiler J.,
     Phys.Rev. A, {\bf 34}, (1986), 2427}
\end{description}
\begin{center}
\begin{tabular}[t]{||l|c|c|c||} \hline \hline
 \em{Series}         & $1/e$   &    $1/10$ &   $0$     \\ \hline \hline
 \em{original}       &   65    &      85   &    92     \\ \hline
 \em{filtered}       &    6    &    12-18  &  22-185   \\ \hline
 \em{smoothed}       &   66    &      85   &    92     \\ \hline
 \em{spring}         &    9    &      17   &    21     \\ \hline
 \em{summer}         &    4    &      13   &    32     \\ \hline
 \em{autumn}         &   10    &      17   &    21     \\ \hline
 \em{winter}         &    7    &      25   &    52     \\ \hline
 \em{differenced}    &    1    &       1   &     1     \\ \hline \hline
\end{tabular}
\end{center}
\PopisTab{1}{The lag (day) at which the autocorrelation function attains the values of
$1/e$, $1/10$ and $0$, for selected series.}
\begin{center}
\begin{tabular}[b]{||l|c|c|c c|c||} \hline \hline
\em SERIES    &   N   & scaling region &   &   $d_f$         & correlation coefficient \\ \hline \hline
\em original  & 78527 &       (1)      & ? & 17.5$\pm$0.3    & 0.99929 \\
              &       &       (2)      &   &1.6$\pm$0.005    & 0.99998 \\ \hline
\em filtered  & 78527 &       (1)      & ? & 21.4$\pm$0.4    & 0.99937 \\ \hline
\em smoothed  & 78523 &       (1)      & ? & 16.2$\pm$0.3    & 0.99934 \\
              &       &       (2)      &   &  1.5$\pm$0.004  & 0.99999 \\
              &       &       (3)      &   &  14.8-17.7      & 0.98393 \\
              &       &       (4)      &   &   1.5-1.7       & 0.99941 \\ \hline
\em spring    & 19780 &       (1)      & ? & 16.4$\pm$0.3    & 0.99922 \\
              &       &       (2)      &   &  4.5$\pm$0.1    & 0.99909 \\ \hline
\em summer    & 19780 &       (1)      & ? & 20.3$\pm$0.4    & 0.99941 \\ \hline
\em autumn    & 19565 &       (1)      &   & 15.9$\pm$0.3    & 0.99914 \\
              &       &       (2)      &   &  3.8$\pm$0.01   & 0.99996 \\ \hline
\em winter    & 19402 &       (1)      & ? & 17.7$\pm$0.3   & 0.99957 \\ \hline
\em differenced & 78526 &              &   & 20.5-26.0       & 0.99989 \\ \hline \hline
\end{tabular}
\end{center}
\PopisTab{2}{The estimates of the fractal dimension $d_f$ for analysed series.
Error represents the 90\,\% confidence limits of the least-squares fit.
The question mark indicates uncertainty of saturation.}
%
\begin{center}
\begin{tabular}{||l|c|c|c|c|c||} \hline \hline
\em SERIES   & scaling region &        $K_2$        &      $T_2$     &  c   & correlation coefficient \\ \hline \hline
\em original & (1)            & 0.14658$\pm$0.03229 & 4.7$\pm$0.9    & 1.50 & 0.99138 \\
             & (2)            & 0.00535$\pm$0.00221 & 129.5$\pm$37.8 & 1.02 & 0.98945 \\ \hline
\em filtered & (1)            & 0.11758$\pm$0.02162 & 5.9$\pm$0.9    & 1.10 & 0.99539 \\ \hline
\em smoothed & (1)            & 0.10909$\pm$0.00501 & 6.4$\pm$0.3    & 1.13 & 0.99919 \\
             & (2)            & 0.00240$\pm$0.00053 & 288.3$\pm$51.7 & 0.97 & 0.99960 \\ \hline
\em spring   & (1)            & 0.17358$\pm$0.01687 & 4.0$\pm$0.4    & 1.61 & 0.99272 \\ \hline
\em summer   & (1)            & 0.14664$\pm$0.01400 & 4.7$\pm$0.4    & 1.34 & 0.99850 \\ \hline
\em autumn   & (1)            & 0.19495$\pm$0.02143 & 3.6$\pm$0.4    & 1.85 & 0.99532 \\
             & (2)            & 0.00892$\pm$0.00168 & 77.7$\pm$12.4  & 1.15 & 0.99651 \\ \hline
\em winter   & (1)            & 0.14717$\pm$0.02322 & 4.7$\pm$0.6    & 1.45 & 0.99566 \\ \hline \hline
\end{tabular}
\end{center}
\PopisTab{3}{\hbox{The estimates of the $K_2$-entropy (1/day) and the error-doubling time
$T_2$ (day) for analysed series.} $c$ is coefficient used for regression.
\hbox{Error represents the 90\,\% confidence limits of the least-squares fit.}}
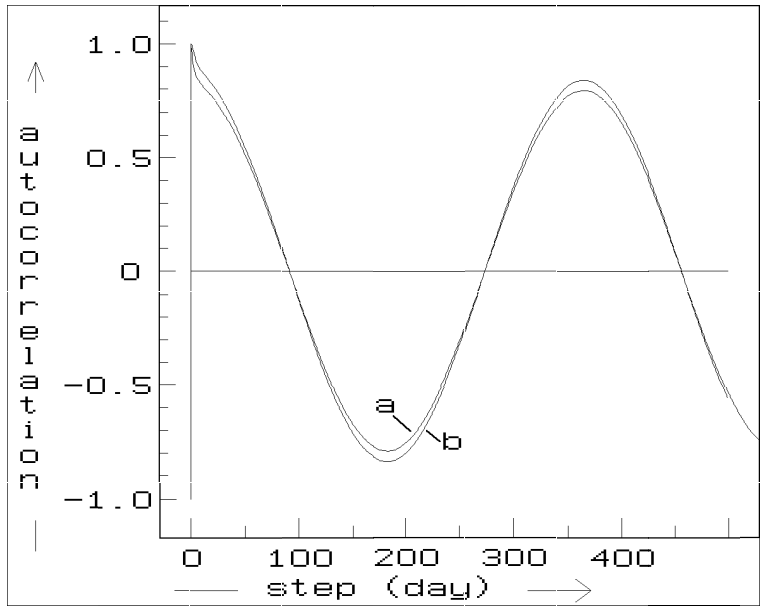
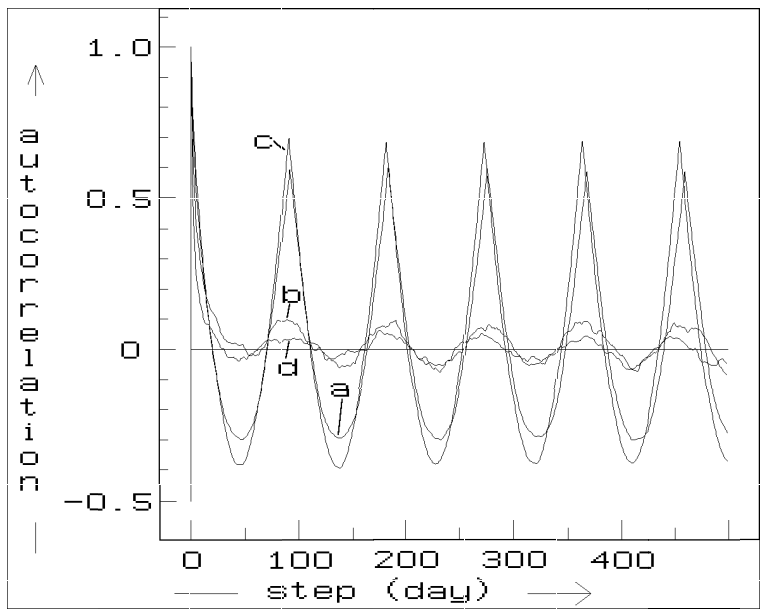
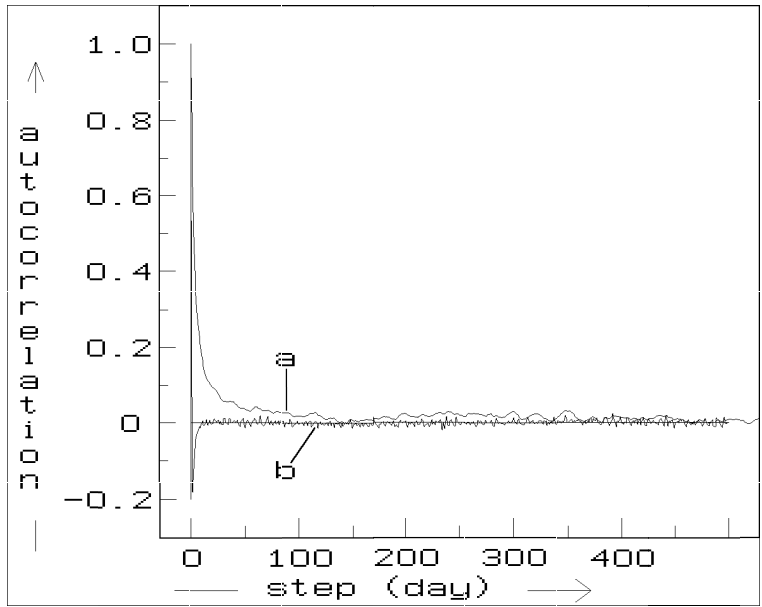
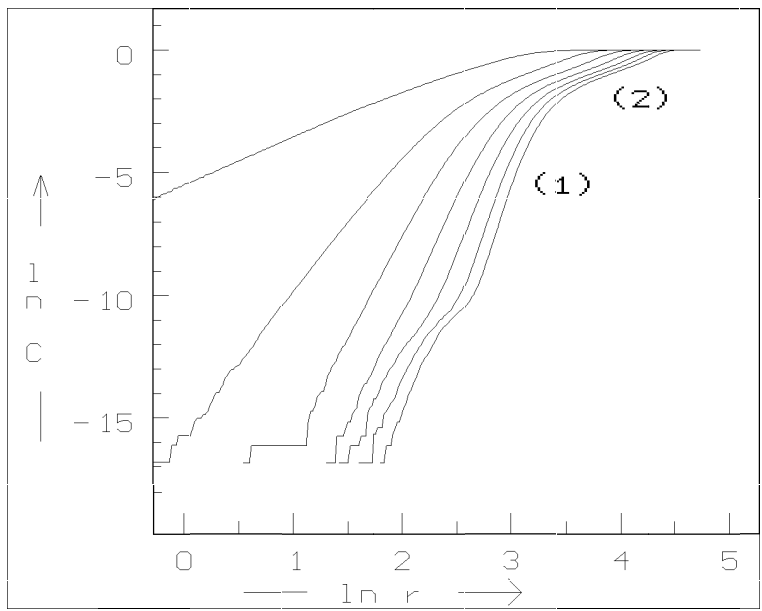
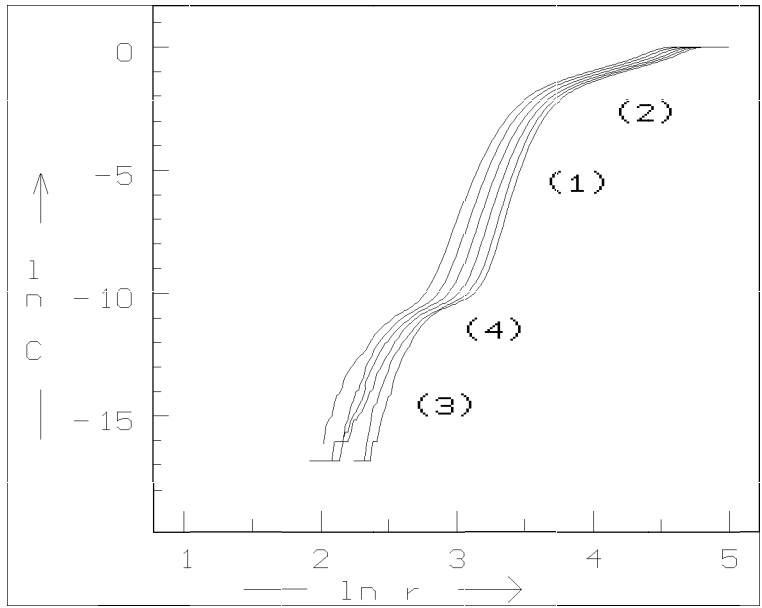
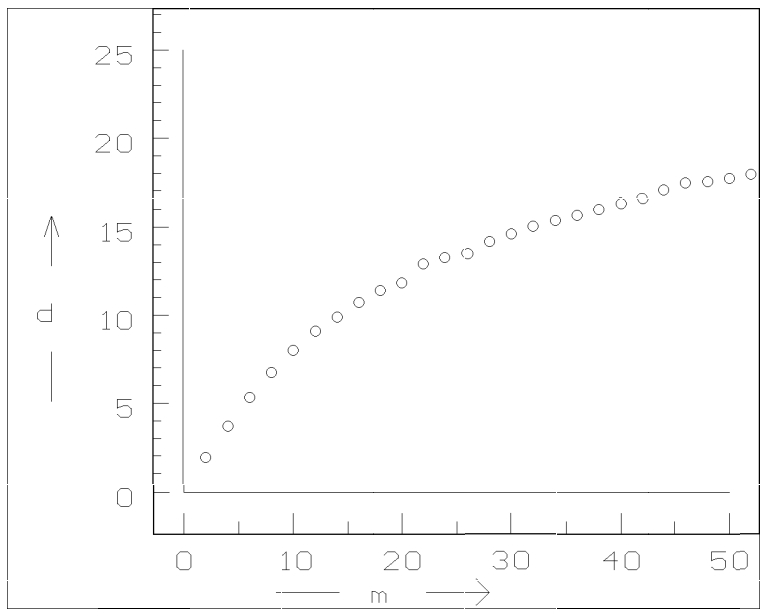
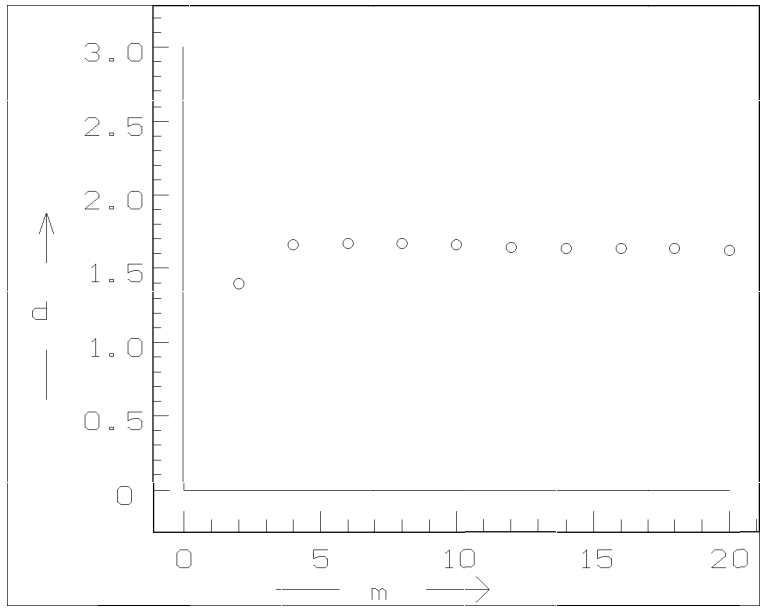
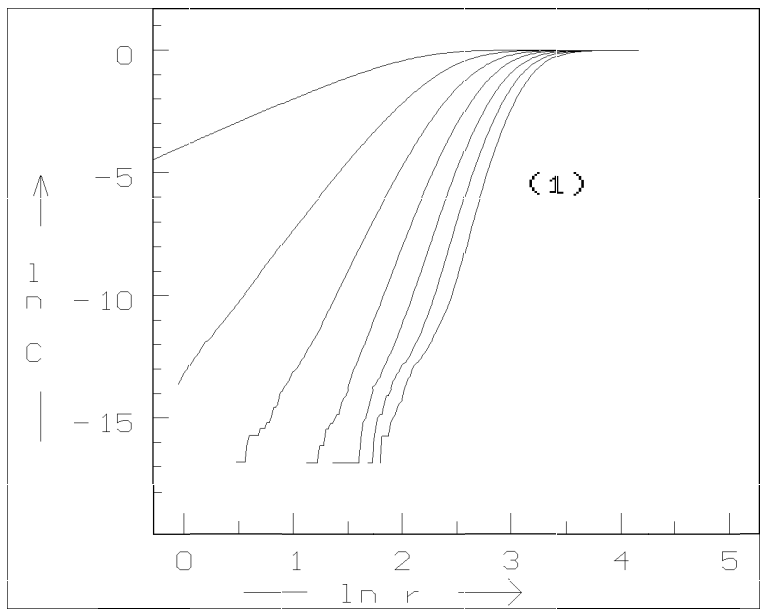
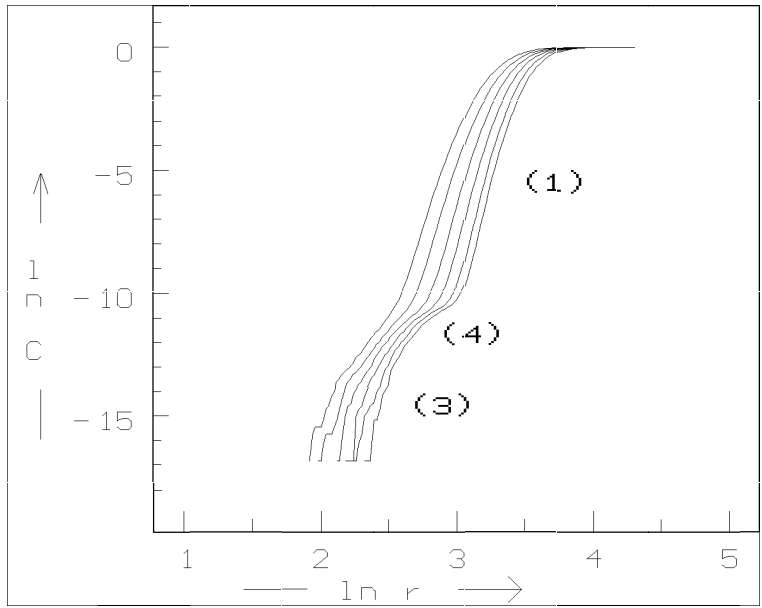
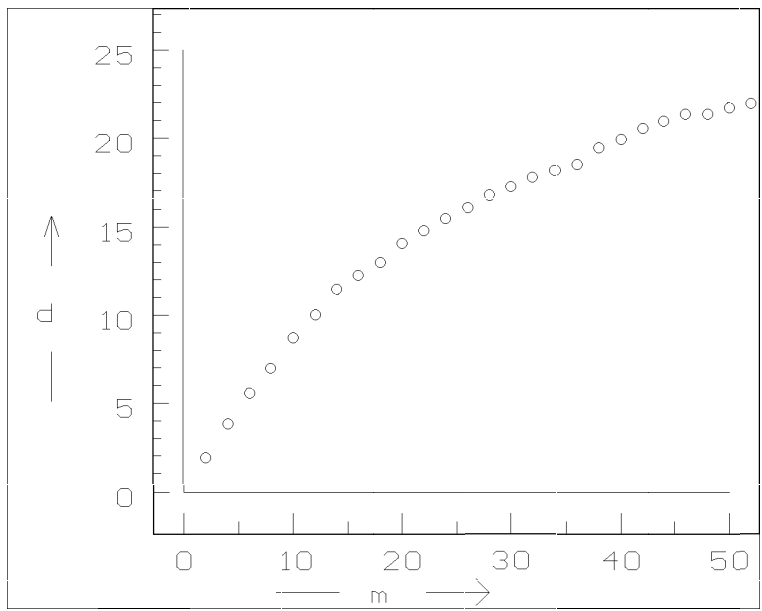
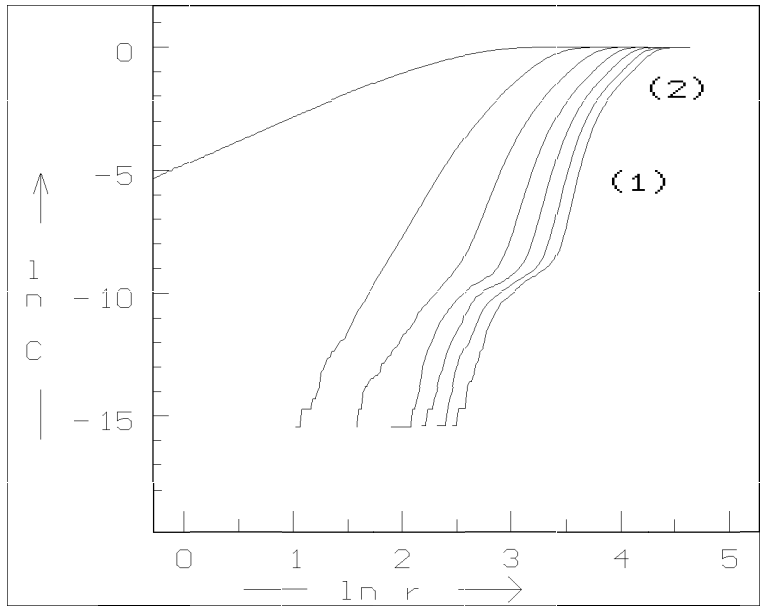
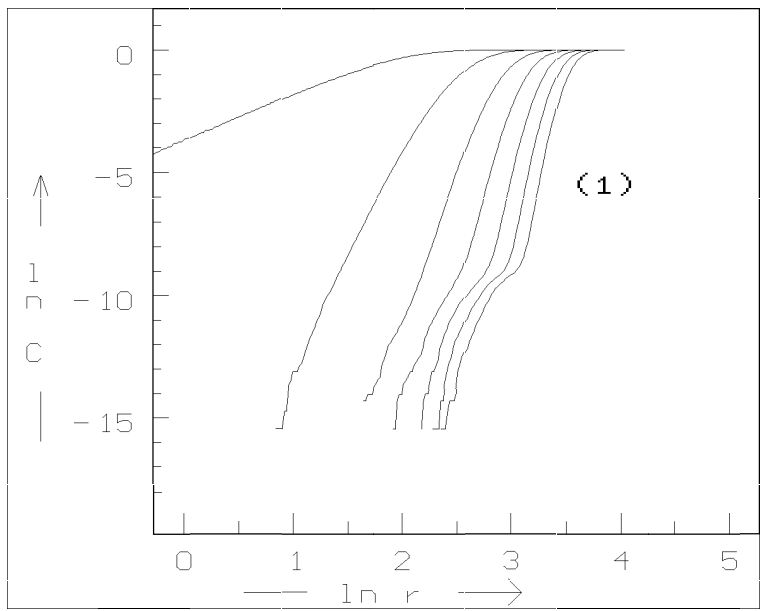
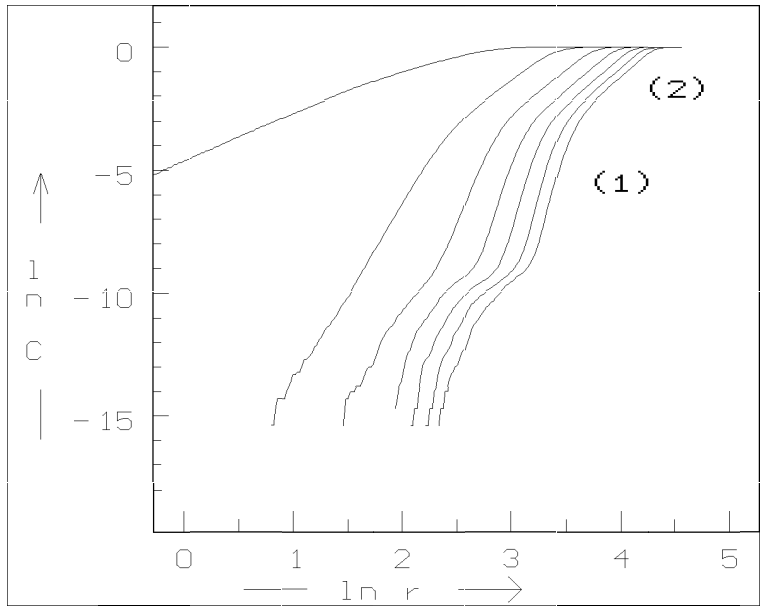
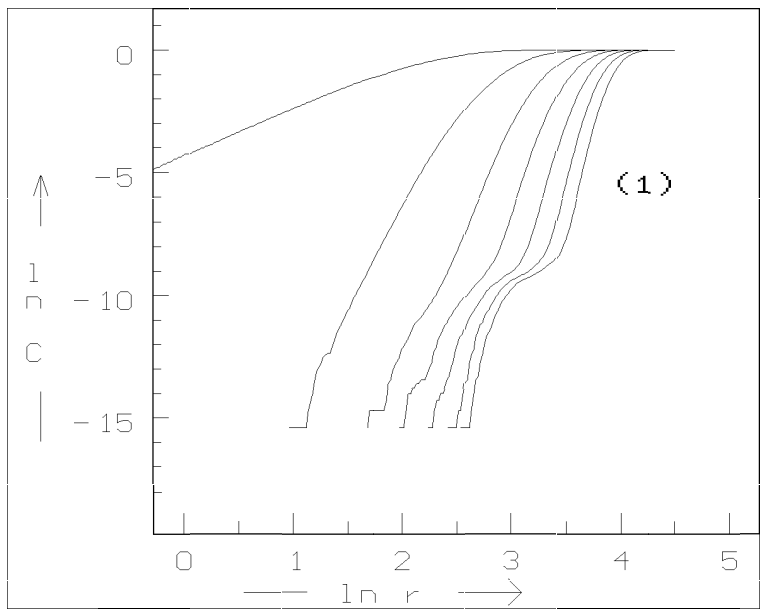
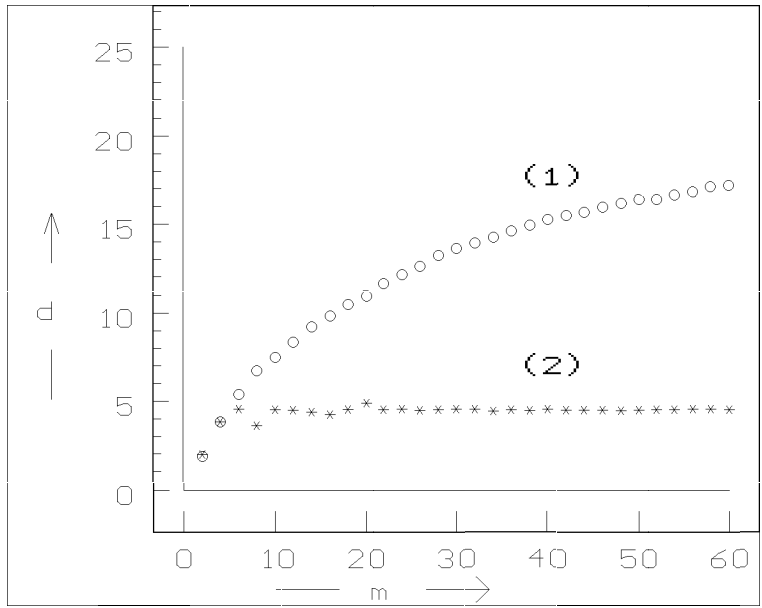
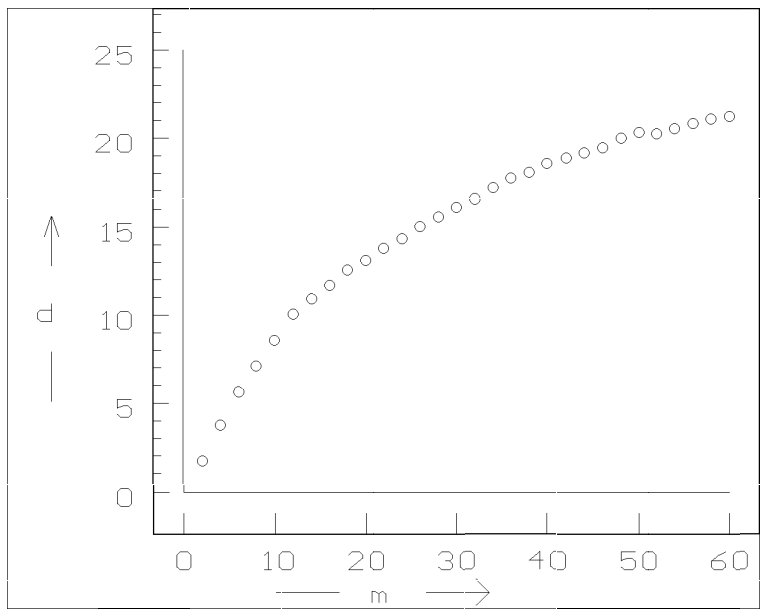
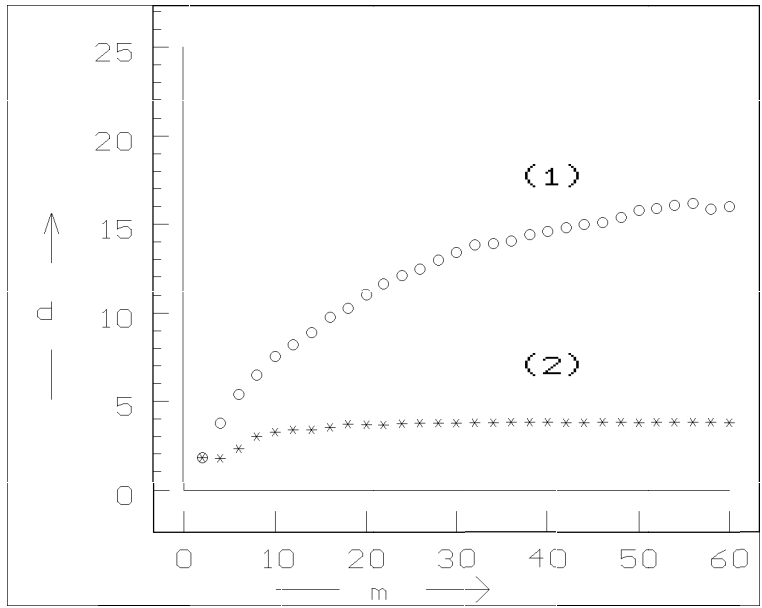
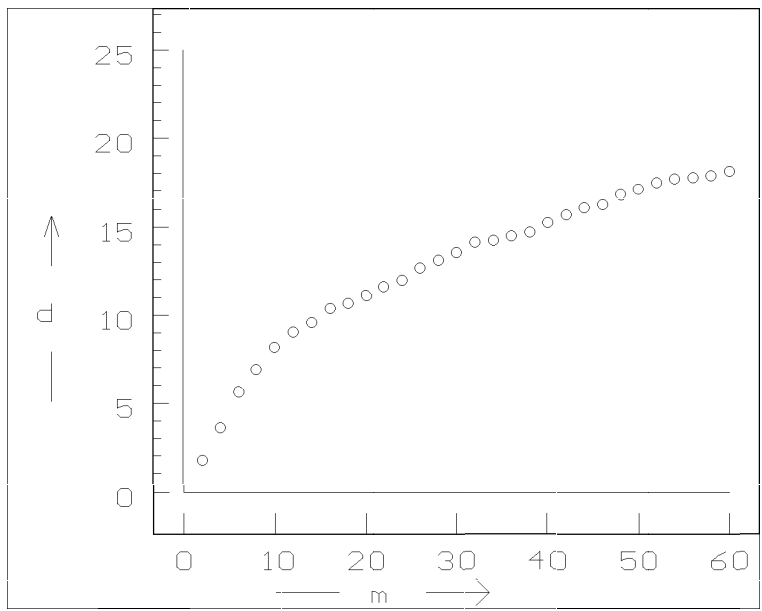
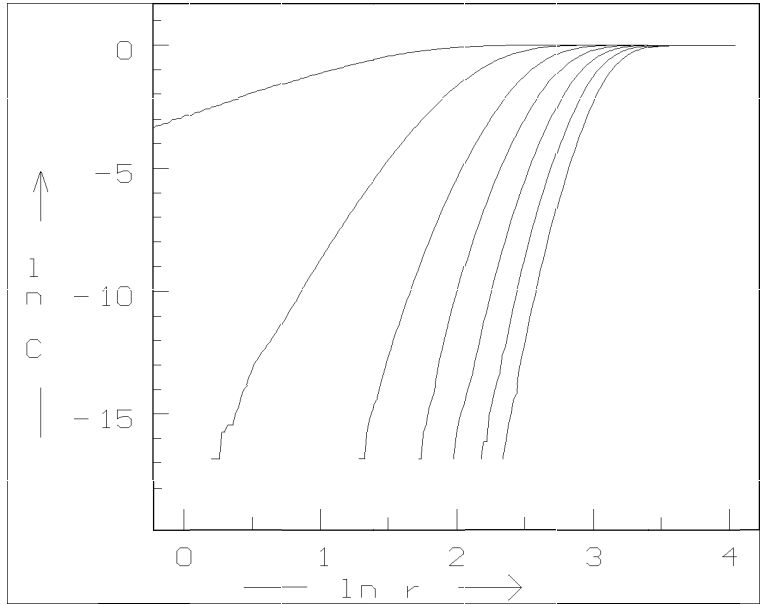
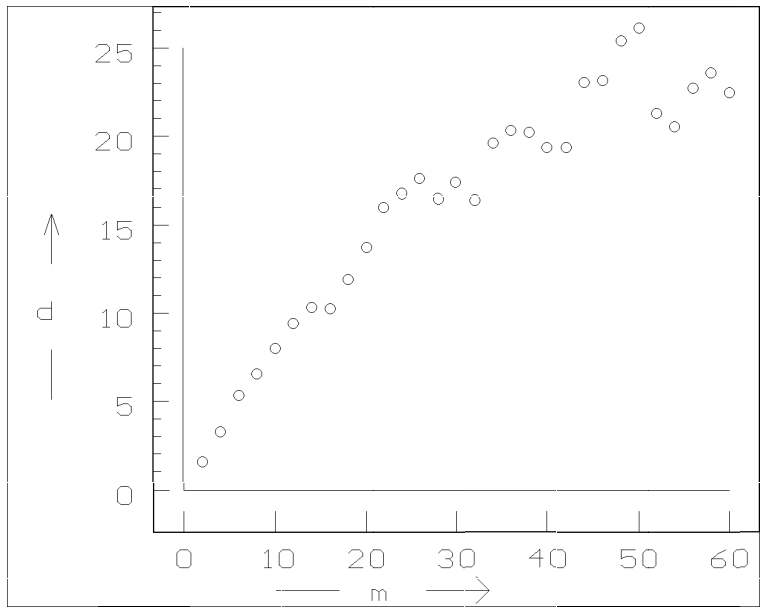
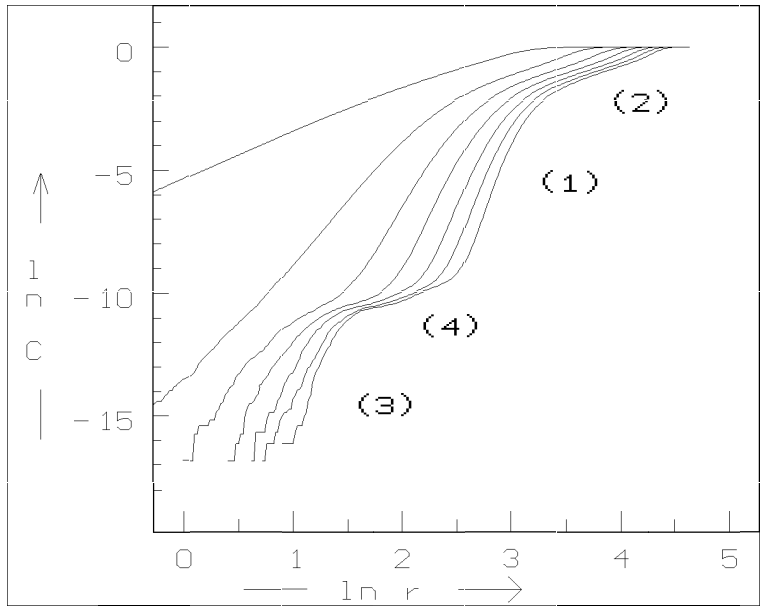
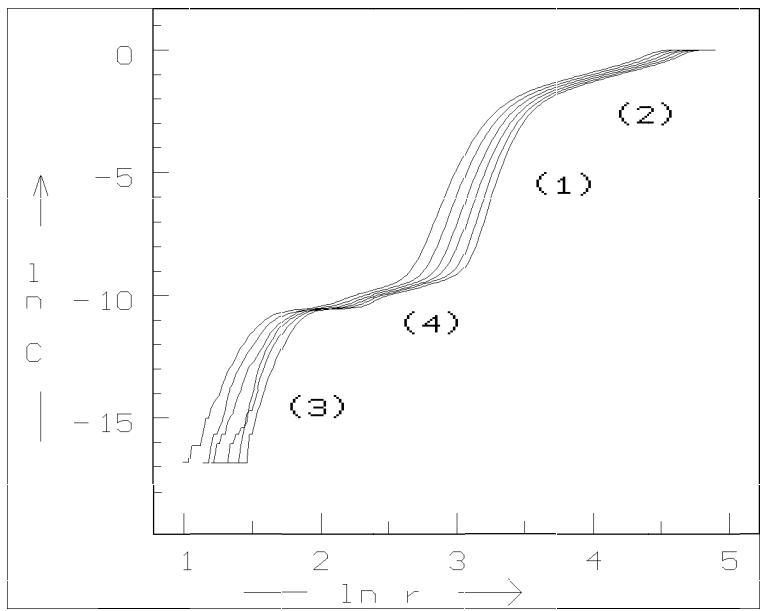
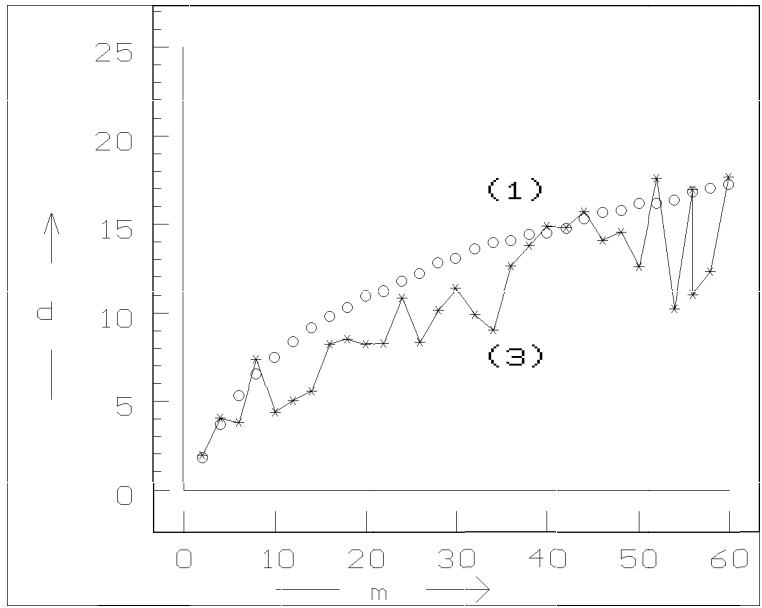
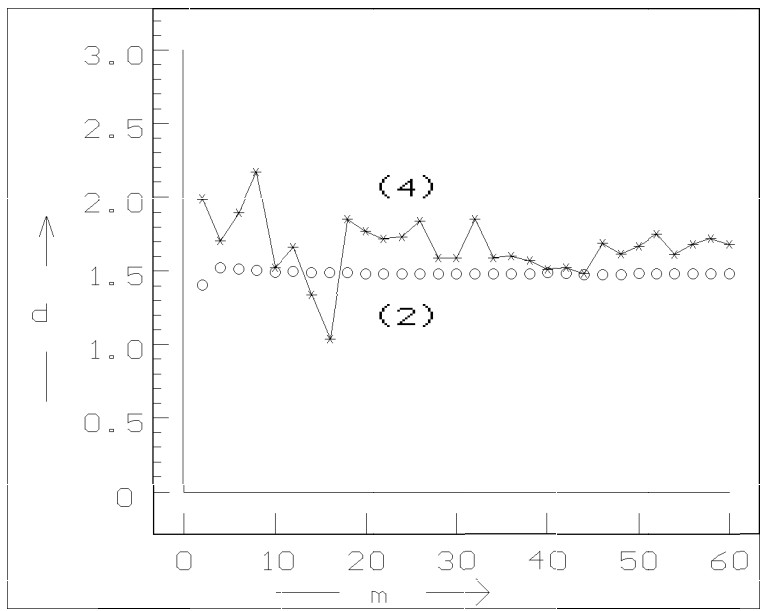
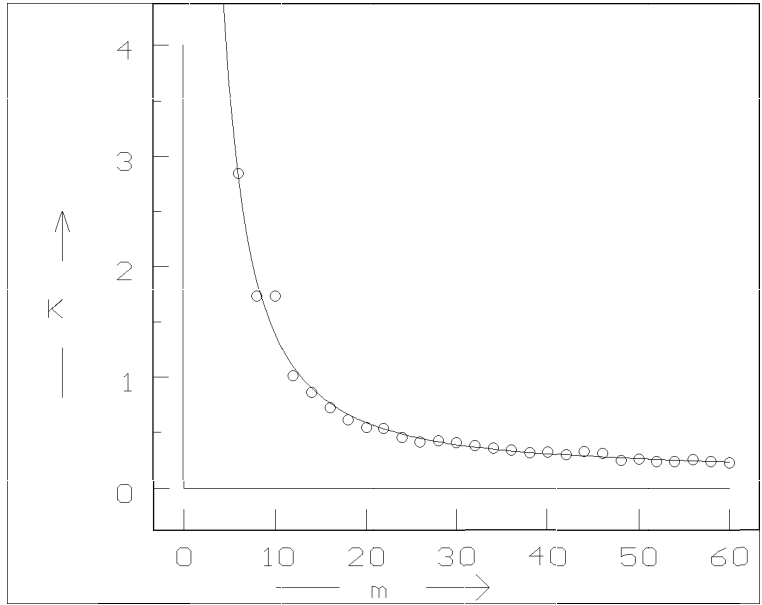
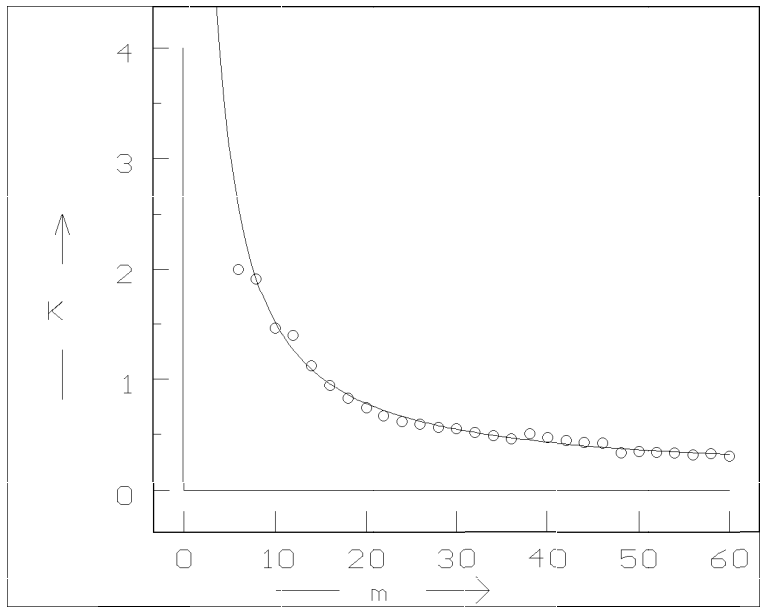
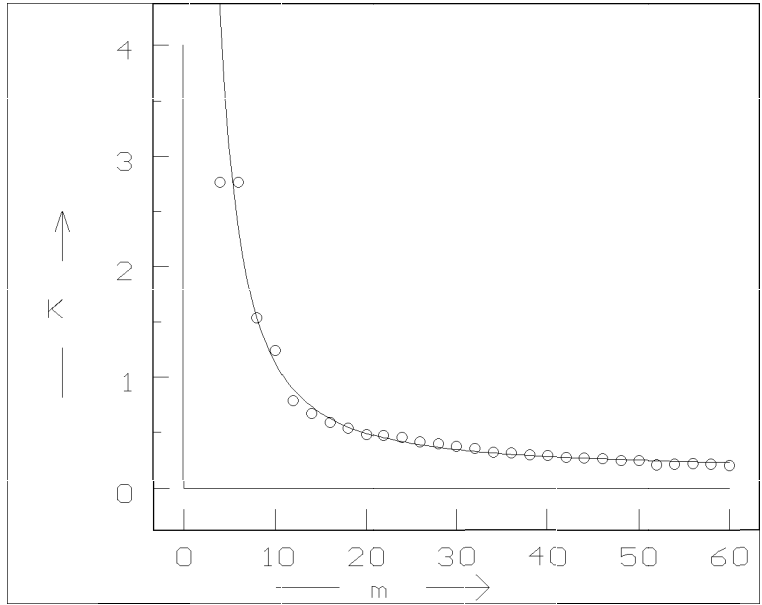
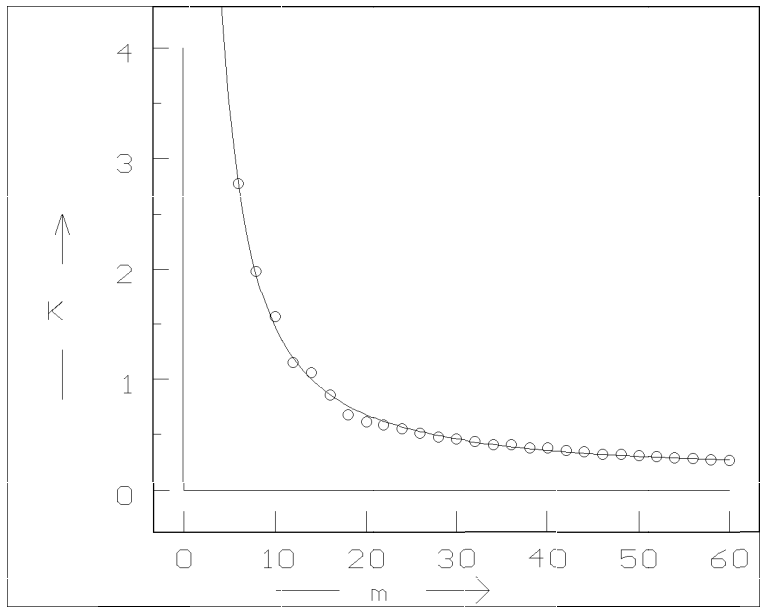
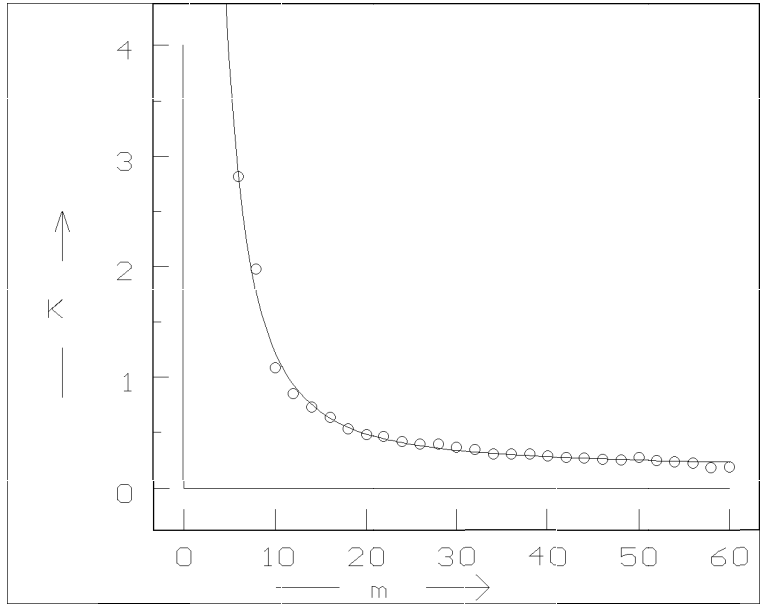
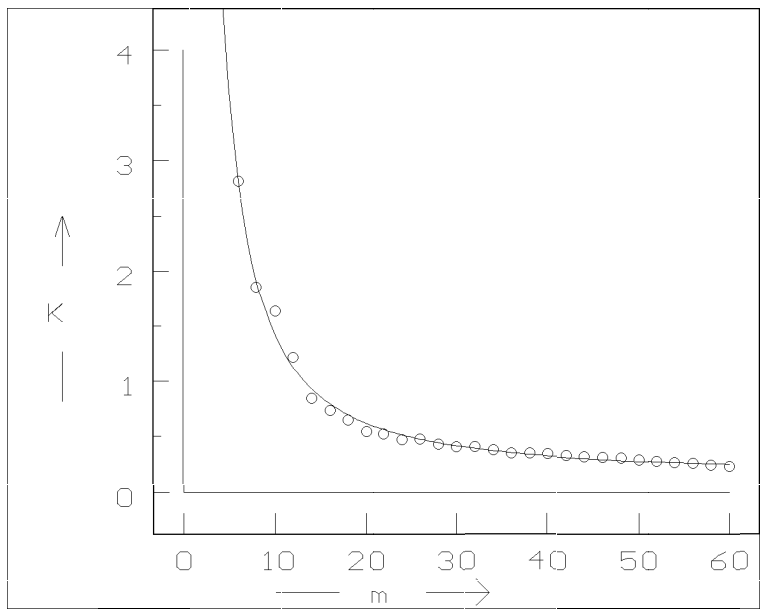
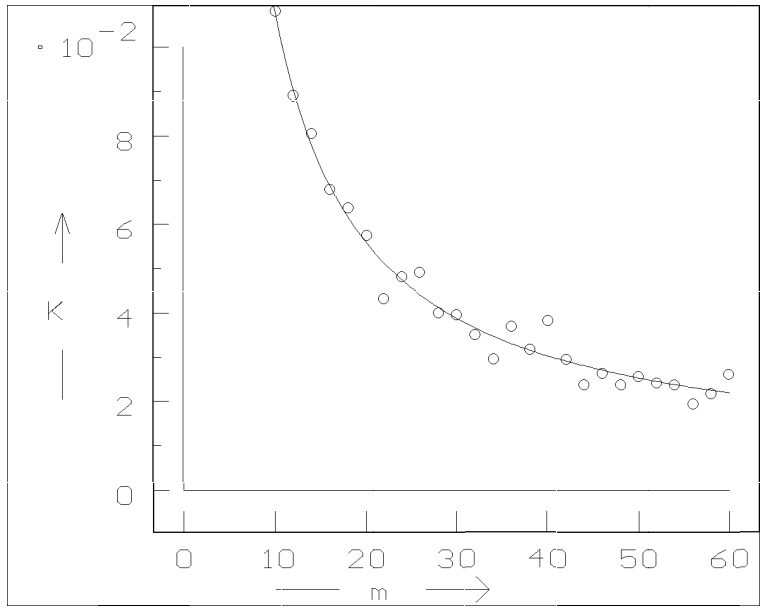
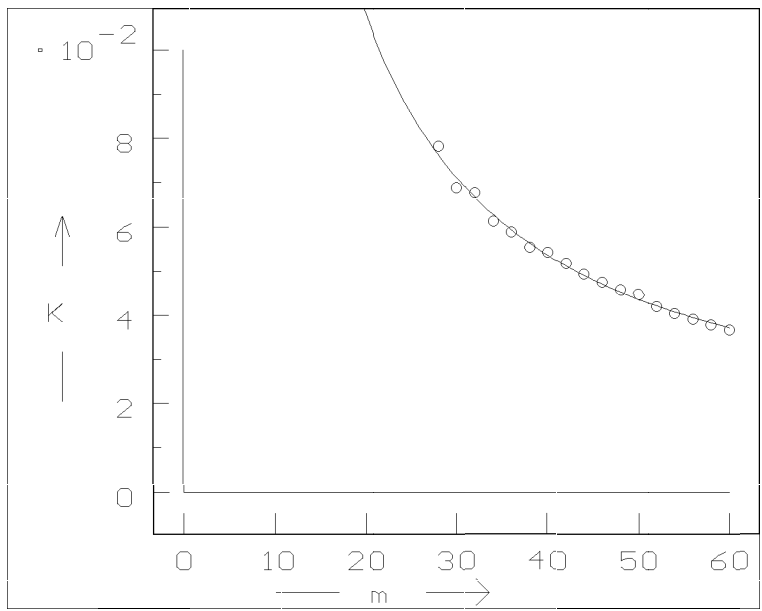
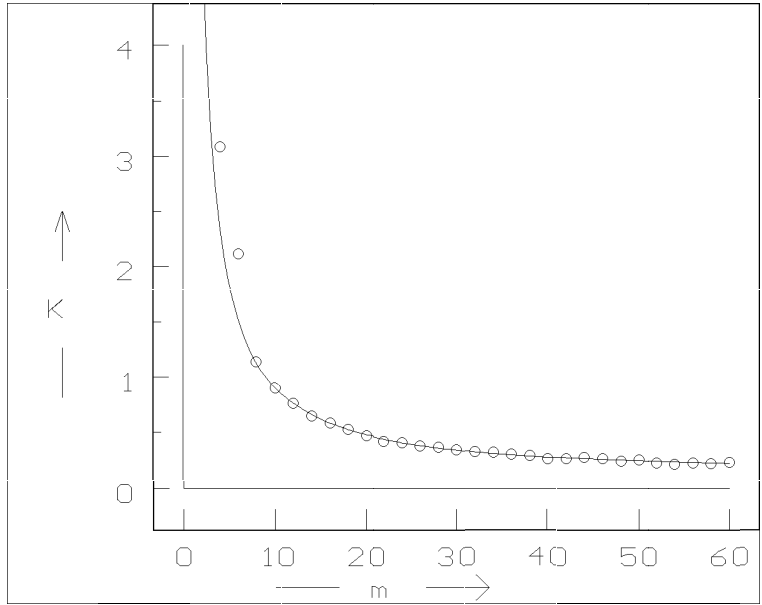
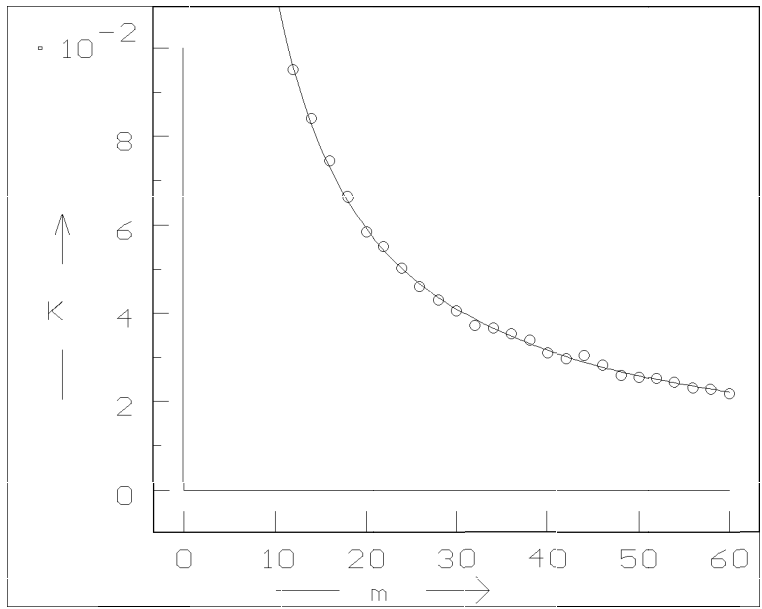
\end{document}